\documentclass[aps,prl,twocolumn]{revtex4}
\usepackage{amsmath,graphicx}

\begin{document}

\title{Optical Tamm States in One-Dimensional Magnetophotonic Structures}
\author{T. Goto$^1$}
\author{A. V. Dorofeenko$^2$}
\author{A. M. Merzlikin$^2$}
\author{A. V. Baryshev$^1$}\thanks{Permanent address: A. F. Ioffe Physico-Technical Institute, St. Petersburg, Russia.}
\author{A. P. Vinogradov$^2$}
\author{M. Inoue$^1$}
\author{A. A. Lisyansky$^3$}
\author{A. B. Granovsky$^4$}

\address{$^1$Department of Electrical and Electronic Engineering, Toyohashi University of Technology, Hibari-Ga-Oka, Tempaku, Toyohashi 441-8580, Japan\\
$^2$Institute for Theoretical and Applied Electromagnetics, Izhorskaya 13/19, Moscow 125412, Russia\\
$^3$Physics Department, Queens College of the City University of New York, Flushing, NY 11367\\
$^4$Faculty of Physics, Moscow State University, Leninskie Gory, Moscow 119992, Russia}

\begin{abstract}
We demonstrate the existence of a spectrally narrow localized surface state, the so-called optical Tamm state, at the interface between one-dimensional magnetophotonic and nonmagnetic photonic crystals. The state is spectrally located inside the photonic band gaps of each of the photonic crystals comprising this magnetophotonic structure. This state is associated with a sharp transmission peak through the sample and is responsible for the substantial enhancement of the Faraday rotation for the corresponding wavelength. The experimental results are in excellent agreement with the theoretical predictions.
\end{abstract}

\maketitle

Within the past two decades progress in nanotechnology has allowed researchers to create complex electromagnetic structures with high accuracy.  Artificial structures with periodically modulated dielectric properties have attracted particular interest. By analogy with crystalline solids, these structures are called photonic crystals (PCs). This analogy has been extremely fruitful since PC optical spectra are in many regards similar to electronic spectra of solids, e.g. a PC spectrum may have a full band gap, may exhibit localization or guiding of light by intentionally introduced defects, also electromagnetic waves may propagate at the PC surface \cite{Yeh1,Yeh2}.

A surface electromagnetic wave may travel along the interface between a single-negative (SNG) \cite{footnote1} and a double positive (DP) media \cite{Landau}. This wave does not propagate into the SNG medium because of its pure imaginary wave number. Due to the total internal reflection, it also does not propagate into the DP medium \cite{footnote2}. This wave is TE polarized in the case of a negative permeability (MNG) medium and TM polarized in the case of a negative permittivity (ENG) medium. In the case of the interface between PC and DP medium, the surface wave can be either TE or TM polarized \cite{Yeh}. Thus, the PC can play a role of either MNG or ENG media. Even more interesting phenomenon is an optical surface mode totally localized on the interface between two different media. This mode does not travel in any direction. As predicted in Ref.~\onlinecite{Engheta}, such a mode may exist at the interface between ENG and MNG media. One may expect that a similar mode may arise when one of the medium is substituted by a PC. Existence of such localized modes was predicted in Refs.~\onlinecite{Villa1,Kavokin,Villa2,Vinogradov,Malkova1,Malkova2}.

As in the case of surface waves, the localized interface mode does not extend into the medium with negative permittivity or permeability because of its pure imaginary wave number. The mode also does not propagate into the PC because its eigenfrequency falls within the photonic band gap. It arises because in a PC the solutions of the Maxwell equations are represented as Bloch waves so that for the modes within the band gap, the exponential fall off of the eigenfunction is modulated by a periodic variation of the dielectric constant. As a result of this periodic variation, one can simultaneously satisfy the continuity conditions for the tangential components of electric and magnetic fields at the boundary. This surface state on a PC is analogous to the nontraveling electron state on a crystal surface predicted by Tamm in 1932 \cite{Tamm}. Because Tamm states in crystalline solids may exist only on very flat surfaces with roughness of an atomic scale, their experimental verification is extremely difficult, and they were first demonstrated only in 1990 \cite{Ohno}. The optical quality of a PC's surface may be precisely controlled, and according to Ref.~\onlinecite{Vinogradov}, the optical Tamm state (OTS) is robust with respect to small surface roughness. Therefore, PCs present an excellent opportunity for the experimental study of OTSs.

In this Letter we report on the first experimental observation of the nonpropagating, $k_{inplane}=0$, surface optical state at an interface of magnetophotonic and nonmagnetic PCs. The state is characterized by a narrow sharp transmission peak at which the Faraday rotation is substantially enhanced. The wavelength of this state falls within the band gaps of both PCs. In the case of a normal incidence of the electromagnetic wave on the interface, the electrodynamics problems maps exactly into the problem considered by Tamm \cite{Tamm}. Therefore we call the observed state OTS.

The structure used in the experiment was made to satisfy conditions for OTS calculated in Ref.~\onlinecite{Vinogradov}. It was comprised of two adjoining one-dimensional PCs deposited on quartz substrates using ion beam sputtering. The first PC was a dielectric multilayer composed of five pairs of Ta$_2$O$_5$/SiO$_2$ films with a Ta$_2$O$_5$ film at the end. The average thicknesses of films were 94 nm (Ta$_2$O$_5$) and 138 nm (SiO$_2$). The second magnetic PC was formed atop of the first PC. Films of bismuth-substituted yttrium garnet (Bi:YIG) were the magnetic constitutive elements of the second PC comprising five repetitions of the Bi:YIG/SiO$_2$ pair; the average thicknesses of Bi:YIG and SiO$_2$ films were 86 nm and 138 nm, respectively. For the sputtered films, deviations of their thicknesses from the desired values were less than 10\%. Each Bi:YIG film was annealed after its deposition in air at 700$^\circ$C for 15 min. The resultant structure was a quartz substrate/Ta$_2$O$_5$/(SiO$_2$/Ta$_2$O$_5$)$_5$/(Bi:YIG/SiO$_2$)$_5$ multilayer [Fig.~1 (a)]. The parameters of the sample were chosen such that the OTS appeared at a wavelength of 800 nm, where Bi:YIG films exhibit a high transmittance and a satisfactory Faraday rotation.

\begin{figure}
  \includegraphics[width=3in]{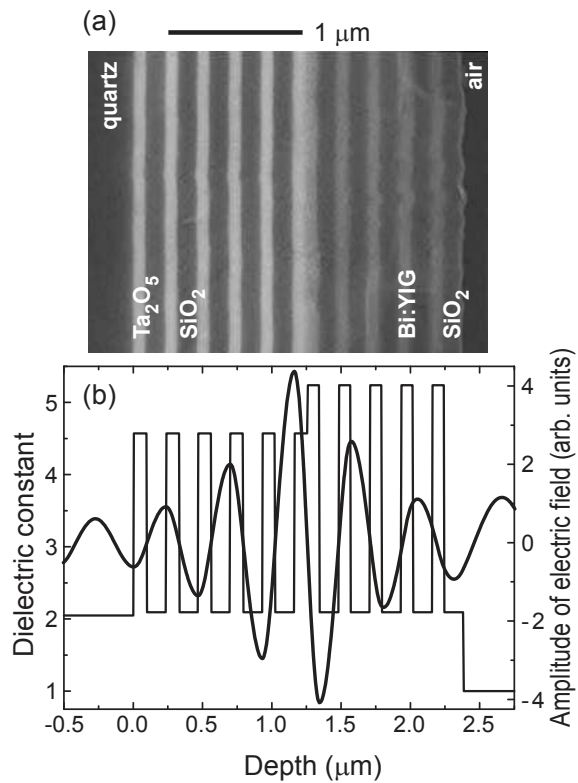}\\
  \caption{SEM image of the system of two adjoining PCs (a); and a sketch of the spatial distribution of dielectric constants and the distribution of the electric field amplitude within the sample (b).}\label{fig:elementary_cell}
\end{figure}

Structural properties of the sample were evaluated using a Jeol JSM-6700 scanning electron microscope (SEM) and Energy Dispersive X-Ray Spectroscopy (EDS). Optical responses were measured by a Shimadzu UV-3100PC spectrophotometer and an apparatus for magneto-optical studies (Neo Ark BH-M600VIR-FKR-TU). According to EDS and magneto-optical measurements, the annealing conditions were sufficient to obtain Bi$_1$Y$_2$Fe$_5$O$_{12}$ single garnet-phase polycrystalline (Bi:YIG) layers with the regular Faraday rotation. Transmission and Faraday rotation spectra were obtained with 5-nm spectral resolution (at normal incidence) for polarized collimated beams; the anglular divergence was less than 1$^\circ$, and the cross-sectional size of the light beam was about 2 mm$^2$. To evaluate the Faraday rotation at extremely low intensities of the transmitted light, i.e., within the photonic band gap, photon-counting mode with long acquisition intervals was used; the accuracy of analysis was 0.02$^\circ$.

Experimental transmission spectra of PC 1, PC 2, and of the combined sample are shown in Fig. 2. One can see that each PC exhibits a band gap in the spectral range of 650-1000 nm, and there is a transmission maximum inside the band gap at 800 nm for the sample. For this wavelength, the distribution of the electric field amplitude within the sample is shown in Fig.~1 (b). The amplitude is remarkably high at the interface between two PCs and falls exponentially away from the interface. Such a distribution confirms the formation of the OTS. Note that all the maxima of the OTS's amplitude are spatially located within (or close to) the Bi:YIG layers. Below, we demonstrate that OTS provides strong optical coupling to the Bi:YIG layers, resulting in an enhanced magneto-optical response.

\begin{figure}
  \includegraphics[width=3in]{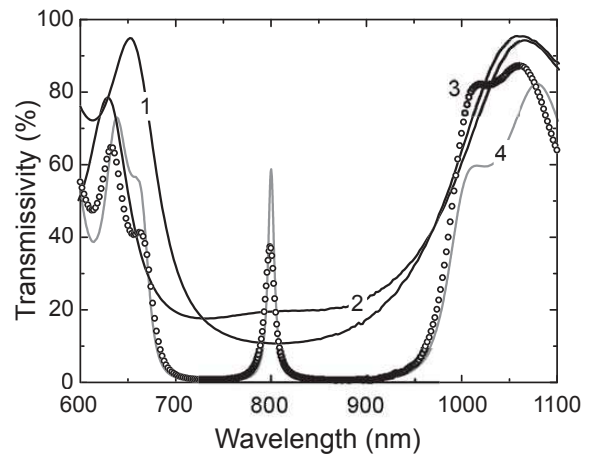}\\
  \caption{Transmission spectra of PC 1 and PC 2 alone are denoted by (1) and (2), respectively. The transmission spectrum of the sample of adjoining PCs is shown by circles (3); its calculated spectrum is given by curve 4. }
\end{figure}

In order to verify that the peak shown in Fig.~2 is due to the OTS, magneto-optical spectra of the reference structure (PC 2) and of the sample were measured. According to the theoretical predictions of Ref.~\onlinecite{Vinogradov}, if one of the adjoining PCs is magnetic, an OTS should cause a substantial enhancement of the Faraday rotation. The results of the measurements are shown in Fig.~3. The Faraday rotation of the PC 2 mainly follows the ordinary response from the Bi:YIG constituent layers. At the transmission peak of 800 nm, the Faraday rotation is $-0.82^\circ$  for the combined sample, which is almost an order of magnitude larger than the value in the PC 2 of $-0.11^\circ$. The peak linewidths of $\sim$10 nm in both the transmission and Faraday rotation spectra are  due to the fluctuations of the layers' thicknesses.

Absorption along with diffuse scattering of the electromagnetic wave due to the roughness of the layer boundaries reduces the effect. This is the reason of the discrepancy of the calculated and measured values of the transmission coefficient shown in Fig. 2. On the other hand, both experiment and theory give practically the same value of the resonance frequency, which weakly depends upon losses.

In sum,both transmission and magneto-optical spectra of the system of adjoining PCs are in excellent agreement with the theoretical calculations also shown in Figs.~2 and 3. These results demonstrate unequivocally the presence of the OTS in the system. As far as we know, unlike propagating surface waves, optical surface states that do not transfer energy in plane were not previously observed.

\begin{figure}
  \includegraphics[width=3in]{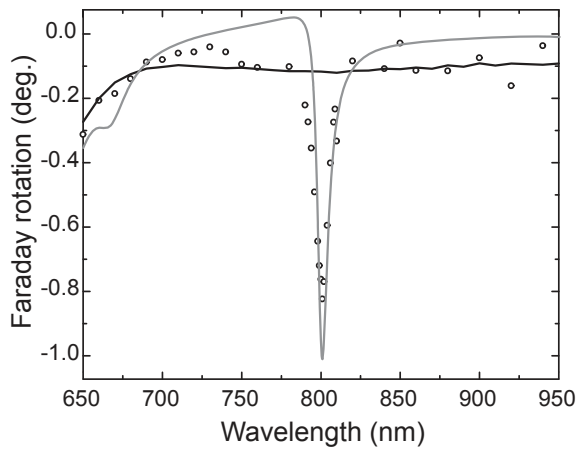}\\
  \caption{Spectra of the Faraday rotation: PC 2 (solid line), the experimental (circles) and calculated (gray line) spectra of the sample.}
\end{figure}

OTSs can be used for magneto-tunable filters theoretically predicted in Ref.~\onlinecite{Merzlikin}. These states can enhance magneto-optical properties of natural materials. Indeed, in the magnetophotonic structures that we discussed, both the Faraday rotation angle, $\theta_F$, and the transmission coefficient, $T$, are large at the resonant frequency. Therefore, the magneto-optical quality factor, $\theta_F/\ln T$, is substantially greater than can be achieved in homogeneous optical material. It is worth noting that structures exhibiting OTSs can be useful for localizing light within any active material used as the constitutive layers of PCs or introduced at the interface between two PCs.

\acknowledgements{This work was supported by RFBR (Grants 06-02-16604, 07-02-91583, 05-02-19886, 06-02-91201, and 08-02-00874-à). AL acknowledges the support by AFSOR, Grant No. FA9550-07-1-0391, and by the CUNY Collaborative grant. The work at TUT was supported in part by Grant-in-Aid for Scientific Research (S) No. 17106004 from Japan Society for the Promotion of Science (JSPS) and the Super Optical Information Memory Project from the Ministry of Education, Culture, Sports, Science and Technology of Japan (MEXT). This work was also supported by Global COE Program ``Frontiers of Intelligent Sensing" from the Ministry of Education, Culture, Sports, Science and Technology of Japan (MEXT).}


\begin{thebibliography}{99}
\bibitem{footnote2} For a surface wave, the in-plane component of the wave vector has a minimum value below which such a wave does not exist \cite{Yeh2}.
\end{thebibliography}
\end{document}